\documentclass[twocolumn]{revtex4}

\newcommand{\beq}{\begin{equation}} 
\newcommand{\eeq}{\end{equation}} 
\newcommand{\beqa}{\begin{eqnarray}} 
\newcommand{\eeqa}{\end{eqnarray}} 
 
\def\<{\langle} 
\def\>{\rangle}

\def\opone{\leavevmode\hbox{\small1\kern-3.8pt\normalsize1}}

\def\Chi{{\chi}}

\begin{document} 
\title{Quantum Gloves} 

\author{D. Collins}
\affiliation{Group of Applied Physics, 20, rue de l'Ecole-de-M\'edecine, 
      CH-1211 Geneva 4, Switzerland, email: d.g.collins.95@cantab.net}
\author{L. Di{\'{o}}si}
\affiliation{Research Insitute for Particle and Nuclear Physics,\\
H-1525, Budapest 114, POB 49, Hungary}
\author{N. Gisin}
\affiliation{Group of Applied Physics, 20, rue de l'Ecole-de-M\'edecine, 
      CH-1211 Geneva 4, Switzerland}
\author{S. Massar}
\affiliation{Physique Th\'{e}orique, {C.P.} 225, Universit\'{e}
Libre de Bruxelles, Boulevard du Triomphe, 1050 Bruxelles,
Belgium} \affiliation{Centre for Quantum Information and
Communication, {C.P.} 165/59, Universit\'{e} Libre de Bruxelles,
Avenue F. D. Roosevelt 50, 1050 Bruxelles, Belgium} 
\author{S. Popescu}
\affiliation{H. H. Wills Physics Laboratory, University of Bristol, Tyndall 
      Avenue, 
      Bristol BS8 1TL, U.K.}
\affiliation{Hewlett-Packard Laboratories, Stoke Gifford, Bristol BS12
  6QZ, U.K.} 
\date{\today} 

\begin{abstract} 
The slogan {\it information is physical} has been so successful 
that it led to some excess. Classical and quantum information can 
be thought of independently of any physical implementation. Pure 
information tasks can be realized using such abstract c- and 
qu-bits, but physical tasks require appropriate physical 
realizations of c- or qu-bits. As illustration we consider the 
problem of communicating chirality. We discuss in detail the physical
resources this necessitates, and introduce the natural 
concept of {\it quantum gloves}, i.e. rotationally invariant 
quantum states that encode as much as possible the concept of 
chirality and nothing more. 
\end{abstract}

\maketitle 


\section{Introduction}\label{int} 

A question which has attracted much attention over the past
years is how to encode a physical quantity into a finite quantum
system, see \cite{H} for many early studies of this question. 
Examples include encoding the time of an event, 
the phase and amplitude of a coherent state,
a direction in
space\cite{H,MP,GP,PS2,PS,CollinsPopescu}, a
reference frame\cite{PS,BBMT}. 
One can
formalize this problem as follows: one party, Alice, has a classical
description of this physical quantity. By this we mean that Alice has
perfect knowledge of the physical quantity. She encodes the
description into the quantum system and sends it, using an ideal quantum
communication channel, to the second party, Bob. Bob then carries out
a measurement. The result of the measurement provides Bob with some
information about the physical quantity. 

An often overlooked aspect  
of this problem is that since Alice wants to
communicate a physical quantity, the nature of the quantum system she
uses to encode the information and the properties of the
communication channel play an essential role in this problem. Thus
there can be essential differences according to whether the particles used 
are bosons or fermions, according to whether
the degrees of freedom are the spin of the particles, their position,
etc.... 
Some discussions of this point can be found in \cite{Rudolph}, 
\cite{CollinsPopescu} and \cite{G} (the latter work is 
based on some of the results presented here, but focuses only on this
aspect).   

In the present work we consider a particularly simple situation
related 
to the question considered in \cite{PS, BBMT} of transmitting a
reference frame. We suppose that Alice only wants to tell Bob the
chirality of her reference frame, ie. whether it is a left or a right
handed reference frame. This question is apparently very simple since
only a binary  quantity must be communicated. But therein lies the
interest of the problem: since there are no real technical difficulties in
this problem one can focus on the essential role of the physical
nature of the communication channel. 
Thus for instance it is impossible to compare chiralities by 
exchanging only classical information, i.e. by sending only 
abstract 0's and 1's. It is  
quite intuitive why this is so: bits measure the quantity of 
information, but have per se no meaning, in particular no meaning 
about geometric and physical concepts. Hence, if our world is 
invariant under {\it left} $\leftrightarrow$ {\it right}, then 
mere information is unable to distinguish between {\it left} and 
{\it right}. In the appendix the relation of this problem to particle
physics is briefly discussed.

Now, information is physical, as Landauer used to 
emphasize and as every physicist knows today. Hence we must 
consider classical bits physically realized in some system. For 
example the bits 0 and 1 could be realized by right-handed and 
left-handed gloves, respectively. It is obvious that such physical 
bits can be used to send chirality information. But bits realized 
by black and white balls couldn't do the job. Furthermore if the
physical bits
can be encoded in a quantum system then the problem is even more
interesting because of the phenomenon of entanglement which allows
Alice to prepare states which have no classical analogue. Indeed by
exploiting this aspect of quantum systems we will show that it is possible to
communicate chirality perfectly. In this context we introduce the natural 
concept of {\it quantum gloves}, i.e. rotationally invariant 
quantum states that encode as much as possible the concept of 
chirality and nothing more.
Furthermore we will show that quantum gloves can be realized in a
very economical way, using very little resources.

The amount of resources required to communicate a physical quantity is
central to our discussion. 
Understanding it increases our understanding of the physical
quantities and how the uncertainty principle
of quantum mechanics puts constraints on the precision with which 
they can be
represented. 
In the case of chirality, since a single bit must be communicated,
Holevo's bound tells us that in principle a single qubit
suffices. However Holevo's bound can generally only be achieved
asymptotically, using block coding. In the present case, since the
system is finite, finding the best encoding is non trivial.
 We shall show that there are
quantum gloves which consists of only a single qubit. However 
a number of tradeoffs between physical resources are still possible,
such as the number of particles which make up the quantum glove, the
volume in space it occupies, the number of qubits communicated, etc...

The question of communicating chirality in the
quantum setting was already introduced in
\cite{Diosi00}. Unfortunately (as was made clear in the final version
of \cite{Diosi00}) the idea presented there does not work as it is
based on the  incorrect assumption that under parity a
spin pointing up in the $\vec n$ direction $|\uparrow_{\vec n}\rangle$
is flipped into a spin
pointing down in the $-\vec n$ direction $|\downarrow_{\vec
  n}\rangle$ for all $\vec n$. 
In fact there are no degrees of freedom which transform
in this way under parity. In particular -see the discussion
below- spin degrees of freedom are invariant under parity.

The paper is organized as follows. In section \ref{II} 
we discuss the problem of
communicating chirality using only classical systems which is of
interest in itself and sets the stage for the quantum problem.
Then we present in section \ref{III} 
a first example of quantum gloves, discussing in
detail the resources required to realize them. In section \ref{IV} we
show that many
different kinds of quantum gloves can be constructed, depending on the
resources used.
In section \ref{VI} 
a unified approach is developed based on the
properties of the {\it chirality operator}, the operator which one
must measure to determine the chirality of one's reference frame. 
We summarize our results in the conclusion. 

\section{Classical Gloves}\label{II}

Before turning to the problem of quantum gloves, let us consider the
simpler problem of classical gloves, ie. classical systems that can
encode chirality. One possibility is of course for Alice to send Bob 
an orthonormal
frame, represented for instance by three orthogonal vectors labeled from one to
three. These vectors could for instance be realized by having Alice
send Bob arrows, labeled from one to three.

On the other hand it is impossible to communicate chirality
using axial vectors only.
An axial vector can for instance be realized physically by a
rotating disc. The axial vector is the angular momentum of the
disc. However if the disc is completely symmetric (and therefore
contains no other directional 
information than its angular momentum), then under
inversion around its center, the spinning disc stays invariant. 
 This means that it is impossible to encode chirality
in one or many axial vectors, ie. in one or many spinning discs, since
under parity the spinning discs stay invariant. 

However it is interesting that one can encode chirality using one
axial vector and one normal vector. Suppose Alice prepares the axial vector
and the normal vector both pointing in the same direction 
and sends them to Bob. For instance this could be realized by a
spinning disc with UP written on one face and DOWN on the other. Then
the axial vector is the angular momentum of the disc, and the vector
is aligned with the axis of the disc and goes from the DOWN face to
the UP face. This disc is no longer invariant under inversion and can
be used to encode the chirality of Alice's reference frame. Indeed if Bob
has opposite chirality he will find that the angular momentum and the
vector pointing from DOWN to UP are opposite.

An alternative 
way of presenting the same thing is  to suppose that 
Alice prepares a spinning disk of angular
momentum $\vec j = (j_x,j_y,j_z)$ and suppose Bob uses a reference
frame inverted about the origin. Then Bob will say that the angular
momentum of the spinning disc has exactly the same components $(j_x,
j_y, j_z)$. But if Alice prepares a vector with components $\vec v =
(v_x, v_y, v_z)$ then Bob will describe this
vector as having 
components $(-v_x, -v_y, -v_z)$. The sign of the scalar product $\vec
v \cdot \vec j$ can thus encode the chirality of the reference frame. 

Note that all these methods are rather uneconomical, and are far from
what we call a perfect glove. Indeed by sending Bob three vectors (her
reference frame), Alice provides him with enough information to align
his reference frame with hers, ie. an infinite amount of supplementary
information is transmitted in addition to the chirality. On the other
hand in the example in which Alice sends Bob a marked spinning disc less
information is conveyed. Indeed a single direction is
transmitted. This could be used to align the z axis of Alice and Bob's
reference frames, but the relative rotation around the z axis would be
undefined. In addition information could also be encoded in 
the angle between the vector $\vec v$ and the
axial vector $\vec a$. We do not know whether classical methods more
economical than this are possible.

\section{Quantum Gloves}\label{III}

\subsection{Setting the problem}

We now turn to the main subject of this article, namely the problem 
of describing the  chirality of a reference frame using quantum
particles.
We begin by describing precisely the setup.
 To this end
let us consider the task of Alice and Bob from the point of view of an
external observer. From the point of view of the external observer
there are in fact four different situations according to whether he
has the same chirality as Alice and/or Bob. In general he describes
what happens as follows.

First Alice
prepares a quantum state $|G\rangle$ which encodes the chirality 
of her reference frame.
If Alice has the same chirality as the external observer, she will
prepare the state $|G^+\rangle$ whereas if she has the opposite
chiraltiy she  prepare the
state $|G^-\rangle=P|G^+\rangle$ where $P$ is the parity
operator. Obviously, in order for Alice to perfectly encode her chirality in the
quantum state the states $|G^\pm\rangle$ must be orthogonal:
$$
\langle G^-|G^+\rangle= 0\ .
$$

Alice then sends the quantum glove to Bob who measures a
chirality operator of which $|G^\pm\rangle$ are two eigenstates with
different  
eigenvalues. More precisely if Bob has the same chirality as the
external observer he will measure the operator
$\Chi$ whereas if he has the opposite chirality as the external
observer he will
measure $P\Chi P$. 

Thus we can describe the quantum gloves in two ways. First we can
consider the quantum
state prepared by Alice $|G^\pm\rangle$ 
and how they transform one into the other under
parity. The second, more
abstract, approach is to consider  
the chirality operator measured by Bob $\Chi$ and how it transforms under
parity $\Chi \to P \Chi P$. 
We will use
both approaches below.

Note that throughout this article we take the parity
operator $P$ to be the unitary operator that realizes 
inversion around the origin. It acts on
vectors as $P \vec v = - \vec v$. The parity operator leaves axial
vectors unchanged (it leaves spinning discs unchanged), 
hence it also leaves spin degrees of freedom (for
instance the spin of an electron)
unchanged.
Acting with the parity operator
twice always yields the identity: $P^2 = I$.

\subsection{First example and general discussion}

The 
nature of the physical resources used to encode chirality plays
an essential role. Indeed Alice cannot use spin
degrees of freedom alone to solve this problem since they are axial
vectors. 
On the
other hand Alice can use the relative positions of particles. Indeed
the relative position of two distinguishable particles, say 
a proton and an electron, can describe a vector. This is the vector
going from the proton to the electron. Under parity (inversion around
the position of the proton) the vector will flip to the opposite vector. Thus
using the relative position of four distinguishable particles one can
describe a reference frame (one at the origin, the other three along
the three axes). 

It should therefore come as no surprise that one can
construct quantum gloves using the relative position of four
particles. What is more interesting is that 
it can be done perfectly (by this we mean that Bob will be certain of
the chirality of Alice's reference frame) using only a small Hilbert
space (effectively Alice only sends a single qubit) in a way
which conveys no information about the orientation of Alice's
reference frame: the quantum glove states are invariant under
rotation.

Another surprising aspect is that with
minor modifications these perfect quantum gloves can be realized with
only two kinds of particles: we need one reference particle (say a proton) to
indicate the origin of the coordinate system, and three other
indistinguishable particles (say electrons). The positions of
the three indistinguishable particles with respect to each other and
with respect to the reference particle encodes the chirality of the
reference frame. 
The restriction that one of the particles is different from the others
can in fact also be dropped, although we do not know whether perfect gloves
are possible in this case.

In section \ref{IV} we will  further generalise this construction and
show that perfect quantum gloves can be 
realized with only three particles. One, the proton, indicates
the origin of the coordinate system, and the other two can be
indistinguishable. But in this case Alice must send more
than a single qubit to Bob. The extra information 
can be used to
convey some information about the orientation of her reference
frame in addition to its chirality. We also show that 
perfect quantum gloves can be realized
by using the relative position of two particles and spin degrees of
freedom and that imperfect quantum gloves can be realized using the
relative position of two particles only.

We now describe how to construct quantum gloves 
involving four particles. We take as
variables the position of the reference particle $\vec x_0$ and vectors
$\vec x_1$, $\vec x_2$, $\vec x_3$ going from the position of the
reference particle
to the positions of
particles 1, 2, 3. We write $\vec x_i = r_i \ \vec n_{\Omega_i}$ where
$r_i=|\vec x_i|$ and $\vec n_\Omega$ is a unit vector pointing in
direction $\Omega$. We can decompose any wave function of the four
particles into a superposition of factorized wave functions of the form:
\begin{equation}
 \varphi(\vec x_{0}) f(r_1, r_2, r_3)
Y_{l_1 m_1}(\Omega_1)Y_{l_2 m_2}(\Omega_2)Y_{l_3 m_3}(\Omega_3)
\end{equation}
where $Y_{lm}$ are the spherical harmonics. The dependence on $\vec
x_{0}$ plays no role in what follows. Momentarily we also drop the
dependence on the radial variables $r_i$. We will come back to them
below.

The parity operator $P$ realizes the reflection
about the position of the reference particle. Thus $P \vec n = - \vec n$. 
It transforms spherical harmonics
according to
\begin{equation}
P Y_{lm} = (-1)^l Y_{lm}\ .
\end{equation}
Thus the product of three spherical harmonics has parity
\begin{equation}
P Y_{l_1 m_1}Y_{l_2 m_2}Y_{l_3 m_3}= (-1)^{l_1 + l_2 + l_3}
Y_{l_1 m_1}Y_{l_2 m_2}Y_{l_3 m_3}\ .
\nonumber
\end{equation}

Let us now consider the following two states:
\begin{enumerate}
\item
 all three particles in S-waves:
\begin{equation}
|S^3\rangle = Y_{00}Y_{00}Y_{00}\ ;\nonumber
\end{equation}
\item
 all three particles in P-waves, in a completely antisymmetric
   state (known as the Aharonov state):
\begin{eqnarray}
|A\rangle&=& \left( Y_{11}Y_{10}Y_{1-1} + Y_{10}Y_{1-1}Y_{11}+
Y_{1-1}Y_{11}Y_{1-0}\right.\nonumber\\
& & \left.- Y_{11}Y_{1-1}Y_{10}-
Y_{1-1}Y_{10}Y_{11}-Y_{10}Y_{11}Y_{1-1}\right)/\sqrt{6}\ . 
\nonumber
\end{eqnarray}
\end{enumerate}
Both states have zero total angular momentum. This implies that they
are invariant under simultaneous rotations of all three
particles. Under parity they transform as $P|S^3\rangle = |S^3\rangle$
and
$P|A\rangle = - | A\rangle$.

The two quantum gloves are defined as
\begin{equation}
|G^{\pm}\rangle = { |S^3\rangle \pm |A\rangle \over \sqrt{2}}\ .
\end{equation}
These states are orthogonal, they are invariant under rotation, and
under parity they transform as
\begin{equation}
P|G^+\rangle = |G^-\rangle \quad , \quad P|G^-\rangle = |G^+\rangle
\ .
\end{equation}
This means that if Alice and Bob have the same chirality, and try to
construct state $G^+$, then they will construct the same state,
independently of the orientation of their reference frames. On the
other hand if they have opposite chirality, then they will construct
opposite states. By sending each other one of 
these states they can unambiguously
learn whether they have the same or opposite chirality:
Alice prepares state
$G^+$ and sends it to Bob. Bob measures in the $G^+$, $G^-$ basis. If
he finds $G^+$ he concludes that they both have the same chirality. If he
finds $G^-$ he concludes that they have opposite chirality. Because the
states are invariant under rotation, this will work independently of
the alignment of their reference frames, and independently of whether
the state was rotated during transmission.

An essential question is to quantify the resources used by Alice to
describe her chirality to Bob, and in particular the amount of
communication used. The simplest answer would be to argue
that since Alice sends Bob one of two states, the total amount of
communication is a single qubit\footnote{In fact one could argue that
  in addition one must count the amount of classical bits required for
  the parties to describe to each other the protocol they will use. 
We will not count this classical information in what follows. The main
reason is because it could be
  exchanged between the parties a very long time before they actually
  send the quantum gloves, and therefore can be dissociated from the
  actual chirality encoding system}. However because the physical nature
of the system used to encode the chirality plays an essential role in
this problem, the answer to this question cannot be reduced to a
single number, and must be investigated in more detail. As an
illustration let us compare the above protocol with a protocol in
which Alice uses 4 particles to encode her classical reference frame:
one particle is at the origin, one particle very far along the +x
direction, one particle very far along the +y direction, one particle
very far along the +z direction. This ``classical reference frame'', 
and the one
obtained by reflection about the origin, are different (the
corresponding quantum states are orthogonal). Thus 
they could be used to encode chirality. Since only two reference frames
are used, this method would also
seem to require only one qubit of communication. However it is clearly
much less economical than the first method. What is the precise origin
of the difference?

A first important point is to consider the possibility that 
during transmission from Alice
to Bob the quantum glove undergoes a random rotation. Because of
this random rotation, Alice cannot send Bob any information about the
relative orientation of their reference frames. But she can still tell
him about the relative chirality of their reference frames (since parity
and rotations commute). By carrying
out this random rotation, one sees an essential difference between the
quantum gloves $|G^\pm\rangle$ and the classical reference
frame. Indeed the quantum gloves are invariant under rotation, hence
the entropy, when the states are randomly rotated, stays 1 qubit. On
the other hand the entropy of the 
classical reference frame, when randomly rotated, becomes
infinite: the classical reference frame not only encodes the chirality,
but also an infinite amount of additional information about relative
orientation. 

(Note that the position $\vec x_0$ of the quantum glove could in principle
also be used to transmit information, this is obviously irrelevant to
the present problem. We can take for instance the position to be in a
pure state $\psi(\vec x_0)$ on which both Alice and Bob agree, in which
case no information can be transmitted in this way).

There is another interesting way to compare the resources used by
different quantum gloves. This is the volume they occupy in
space. Indeed if we suppose that the particles use hydrogen like
orbitals, then the lower the angular momentum of the particles, the
closer they can lie to the origin. Thus $L_{max}$, the largest angular
momentum of the particle, measures how much space they occupy. (For
the quantum gloves described above $L_{max}=1$ whereas for the
classical reference frame $L_{max}=\infty$). Finally the number
and type 
of particles used to realize the quantum gloves is another type of
resource which can be compared (indeed we shall describe below quantum
gloves using less than 4 particles). 

This discussion shows that
there is not a unique parameter which quantifies how much
resources are used to encode the chirality of a reference frame. This is
because encoding chirality cannot be done without reference to the physical
system that is used. Thus whereas the resources used in 
many quantum communication tasks can
be quantified in terms of universal units such as bits, qubits,
ebits, in the case of physical quantities protocols will be
inequivalent when different physical systems are used to encode the
same physical quantity.

\section{Many other quantum gloves}\label{IV}

\subsection{Quantum gloves made from indistinguishable particles}

We now go back to describing different kinds of quantum gloves.
As a first extension of the above protocol, let us note that it required the
four particles sent by Alice to be distinguishable.
However it can easily be extended to the case where some or all of the 
particles are
indistinguishable. But now one needs to take care that the global wave
function 
is symmetric or antisymmetric according to whether one is
dealing with bosons or fermions. 

 As illustration we take the reference particle to be distinguishable
from the other three particles which are taken to be indistinguishable
fermions.  
Define
$f^{symm(anti)}(r_1,r_2,r_3)$ to be symmetric (antisymmetric)
functions of the radial coordinates $r_1, r_2, r_3$ respectively. Then
the global wave function of the quantum gloves can be taken to be
\begin{equation}
|G^{\pm}\rangle=\varphi(\vec x_0) \left(
f^{anti}|S^3\rangle \pm f^{symm} |A\rangle \right)/\sqrt{2} \ .
\end{equation}
The case of three bosons is similar except that $f^{symm}$ and $f^{anti}$
are interchanged. We do not know whether it is possible to realize
quantum gloves with only indistinguishable particles (here 
the reference particle $\vec x_0$ is different from the other three).
Note that because the radial wave functions must be symmetric and
antisymmetric, the particles will occupy a larger volume in space than
in the case of distinguishable particles. Thus one has relaxed one
condition (that the particles be distinguishable), but one has had to
use more of another resource (space) in order to make these quantum
gloves. 

One can in fact take all particles to be indistinguishable. One
possibility is to take one particle as particle 1, two particles close together
as particle 2, three particles close together as particle 3,
etc...We do not know however what is the optimal way of doing this.

\subsection{Quantum gloves made from three particles}

We now show how Alice can encode the chirality of her reference frame in
the relative position of 3 particles, one of which (the proton) 
is distinguishable
from the other two (the electrons). We take as
variables the position of the reference particle $\vec x_0$ and vectors
$\vec x_1=r_1 \vec n_{\Omega_1}$, $\vec x_2 = r_2 \vec n_{\Omega_2}$ 
going from the position of the
reference particle
to the positions of
particles 1 and 2. 
We can decompose wave functions of the three
particles into factorized wave functions of the form:
\begin{equation}
 \varphi(\vec x_{0}) f(r_1, r_2)
Y_{l_1 m_1}(\Omega_1)Y_{l_2 m_2}(\Omega_2)
\label{psi3}
\end{equation}
From now on we drop the
dependence on $\vec x_0$ and on $r_1, r_2$.

The rules of addition of angular momentum imply that all states with
zero total angular momentum are combinations of states of the form
eq. (\ref{psi3}) with $l_1=l_2$. Hence they always have parity $P=+1$
and cannot be used to encode the chirality of a reference frame.
However there exist spaces with total angular momentum $L_{TOT}\neq
0$ of opposite parity. 
For simplicity we consider the case
$L_{TOT}=1$. Thus for instance the states
\begin{eqnarray}
|\alpha_{11}\rangle &=& {Y_{00} Y_{11} + Y_{11} Y_{00} \over \sqrt{2}}
\ , \nonumber\\
|\alpha_{10}\rangle &=& {Y_{00} Y_{10} + Y_{10} Y_{00} \over \sqrt{2}}
\ , \nonumber\\ 
|\alpha_{1-1}\rangle &=& {Y_{00} Y_{1-1} + Y_{1-1} Y_{00} \over \sqrt{2}}
\nonumber
\end{eqnarray}
form a basis of an irreducible representation of the rotation group
with total angular momentum $L_{TOT}=1$ and parity $P=-1$. On the
other hand the states
\begin{eqnarray}
|\beta_{11}\rangle &=& {Y_{10} Y_{11} - Y_{11} Y_{10} \over \sqrt{2}}
\ , \nonumber\\ 
|\beta_{10}\rangle &=& {Y_{1-1} Y_{11} - Y_{11} Y_{1-1} \over \sqrt{2}}
\ , \nonumber\\ 
|\beta_{1-1}\rangle &=& {Y_{1-1} Y_{10} - Y_{10} Y_{1-1} \over \sqrt{2}}
\nonumber
\end{eqnarray}
form a basis of an irreducible representation of the rotation group
with total angular momentum $L_{TOT}=1$ and parity $P=+1$.

The quantum gloves consist of two spaces of dimension 3, each of which
constitutes an irreducible representation of the rotation group with
total angular momentum $L_{TOT}=1$. A basis of this spaces is
\begin{eqnarray}
|G^\pm_{11}\rangle &=& {\alpha_{11} \pm \beta_{11} \over \sqrt{2}}
\ , \nonumber\\
|G^\pm_{10}\rangle &=& {\alpha_{10} \pm \beta_{10} \over \sqrt{2}}
\ , \nonumber\\
|G^\pm_{1-1}\rangle &=& {\alpha_{1-1} \pm \beta_{1-1} \over \sqrt{2}}
\nonumber
\end{eqnarray}
where $|G^{\pm}_{LM}\rangle$  is a quantum glove state with total
angular momentum $L$ and angular momentum along the z axis equal to
$M$. 
We denote by $\Pi_{G^\pm}$ the projectors onto these spaces. Thus
\begin{equation}
\Pi_{G^+} =
|G^+_{11} \rangle \langle G^+_{11} | + 
|G^+_{10} \rangle \langle G^+_{10} | + 
|G^+_{1-1} \rangle \langle G^+_{1-1} | 
\end{equation}
and similarly for $\Pi_{G^-}$. It is immediate to check that $\langle
G^+_{1M} | G^-_{1M'}\rangle = 0$ for all $M$, $M'$. This implies that
the projectors $\Pi_{G^+}$ and $\Pi_{G^-}$ are orthogonal.
Furthermore we note that the projectors $\Pi_{G^\pm}$ are invariant 
under simultaneous rotations of both
particles 1 and 2 around the reference particle. This follows from the
fact that they project onto the spaces spanned by all
the vectors of an irreducible representation of the rotation group.
Finally we note that
under parity these projectors transform as
\begin{equation}
P \Pi_{G^+} P = \Pi_{G^-}
 \quad , \quad P \Pi_{G^-} P = \Pi_{G^+}
\ .
\end{equation}

This means that if Alice and Bob have the same chirality, then their
definitions of $\Pi_{G^\pm}$ will coincide. But if they have opposite
chirality, then what Alice calls $\Pi_{G^+}$, Bob will call $\Pi_{G^-}$,
and similarly what Alice calls $\Pi_{G^-}$, Bob will call $\Pi_{G^+}$.
 The protocol is then similar to the previous case: Alice prepares a state in
$\Pi_{G^+}$ and sends it to Bob. Bob projects the state onto the
$\Pi_{G^\pm}$ spaces. If he finds space $G^+$, he concludes they both
have the same chirality, and if he finds space $G^-$, he concludes that
they have opposite chirality. Note that because the spaces
$\Pi_{G^\pm}$ are invariant under rotation, this will work even if the
state Alice sends undergoes a random rotation during
transmission. Conversely if the state does not undergo any rotation,
Alice can send Bob some information about the orientation of her
reference frame. For instance if she sends Bob the state
$|G^\pm_{11}\rangle$ aligned with her z axis, then by measuring the
state Bob can learn information about the orientation of Alice's
reference frame. If one averages over rotations, then this protocol
uses $1+\ln 3$ qubits of communication, whereas the protocol using 4
particles used only 1 qubit of communication. Once more one sees how
one resource is traded for anther.
The above protocol can be generalized to the case
where particles 1 and 2 are identical exactly as in the case where
four particles were sent.

\subsection{Quantum gloves made from spins and relative positions}

As we mentioned above it
is also possible to encode the chirality of a reference frame using one 
vector  and one axial vector. 
This can be done classically by having Alice prepare a spinning disk of
angular momentum $\vec j$ and a vector $\vec v$. The sign of the
scalar product $\vec j \cdot \vec v$ then encodes the chirality of the
reference frame.

An interesting 
semi-quantum implementation of this construction is for Alice to send Bob
a photon propagating along direction $\vec v$ with right circular
polarization. Bob then measures the photon in the right/left circular
basis. The system sent in this case has both a classical degree of
freedom (the direction of propagation) and a quantum degree of freedom
(the spin of the photon). Hence if one averages the system over the
rotation group, one finds that its entropy becomes infinite: this
system requires an infinite number of qubits.

But this construction also has a purely quantum implementation: 
by using two spin
$1/2$ degrees of freedom and one relative position it is possible to
construct two states of total angular momentum zero (and therefore
invariant under rotation) but of opposite parity:
\begin{eqnarray}
|\alpha\rangle &=& |\mbox{Singlet}\rangle Y_{00}\nonumber\\
|\beta\rangle &=& { |\uparrow\uparrow\rangle Y_{1-1}
- |\mbox{Triplet}\rangle Y_{10}
+ |\downarrow\downarrow\rangle Y_{11} \over \sqrt{3}}
\end{eqnarray}
where $|\mbox{Singlet}\rangle = (|\uparrow\rangle|\downarrow\rangle -
|\downarrow\rangle|\uparrow\rangle)/\sqrt{2}$ and
$|\mbox{Triplet}\rangle = (|\uparrow\rangle|\downarrow\rangle +
|\downarrow\rangle|\uparrow\rangle)/\sqrt{2}$.
We now use the
fact that any wave function $|\psi_{spin}\rangle$  composed only of  
spin degrees of freedom is invariant under parity: $P|\psi_{spin}\rangle =
|\psi_{spin}\rangle$.  This implies that
 $P|\alpha\rangle = + |\alpha\rangle$ and $P|\beta\rangle =
-|\beta\rangle$, hence $|G^{\pm}\rangle = (|\alpha \rangle \pm |\beta
\rangle )/\sqrt{2}$ constitute good quantum gloves. 

With a single spin $1/2$ and the relative position of two
particles one cannot construct a state of total angular momentum
zero. 
However one can construct two spaces of dimension 2, of total angular momentum
$1/2$,  and of opposite parity. 
Bases of these spaces are:
\begin{eqnarray}
|\alpha_{1/2 +1/2}\rangle &=& Y_{00} |\uparrow\rangle\ ,\nonumber\\
|\alpha_{1/2 -1/2}\rangle &=& Y_{00} |\downarrow\rangle\ ;
\end{eqnarray}
and
\begin{eqnarray}
|\beta_{1/2 +1/2}\rangle &=& \frac{Y_{10} |\uparrow\rangle\
- \sqrt{2} Y_{11}|\downarrow\rangle}{\sqrt{3}},\nonumber\\
|\beta_{1/2 -1/2}\rangle &=&  \frac{Y_{10} |\downarrow\rangle\
- \sqrt{2} Y_{1-1}|\uparrow\rangle}{\sqrt{3}} .
\end{eqnarray}

\subsection{Quantum gloves made from relative position of two particles}

Finally let us show that one can construct approximate
quantum gloves using the relative position of two particles only. That
this should be the case can be seen from the example discussed above
of the spinning disk with asymmetric upper and lower sides. Indeed
consider an electron in orbit around a proton. The angular momentum of
the electron defines the axial vector $\vec a$. However the wave
function of the electron need not be symmetric between the upper and
lower sides of the plane of rotation, hence encoding a vector parallel
to $\vec a$. The arguments given above show that this should allow Alice
to encode the chirality of her reference frame. We now show how this can
be done.

Consider the following two orthogonal states:
\begin{eqnarray}
|g_+\rangle &=& { Y_{00} + Y_{01} \over \sqrt{2}}\nonumber\\
|g_-\rangle &=& { Y_{00} - Y_{01} \over \sqrt{2}}\ .
\end{eqnarray}
Under parity they transform as $P |g_+\rangle = |g_-\rangle$ and
$P|g_-\rangle = | g_+\rangle$. Thus they seem good candidates for
quantum gloves. However these are not perfect quantum gloves because
under rotation $|g_+\rangle$ does not stay orthogonal to
$|g_-\rangle$. Thus if Alice and Bob's reference frames are not
aligned, or if the quantum glove undergoes a random rotation during
transmission, then they cannot learn with certainty whether they have
the same chirality. More precisely one computes that
\begin{eqnarray}
\rho_{\pm}&=&\int dR \ U_R |g_\pm\rangle\langle g_\pm|U_R^\dagger\label{RR}\\
&=&
\frac{1}{6}\left(
|Y_{11}\rangle\langle Y_{11}| + 
|Y_{10}\rangle\langle Y_{10}|
|Y_{1-1}\rangle\langle Y_{1-1}|\right)
+\frac{1}{2}|Y_{00}\rangle\langle Y_{00}|\nonumber\\
& &\pm \frac{1}{4}\left(|Y_{00}\rangle\langle Y_{10}|
+|Y_{10}\rangle\langle Y_{00}|\right)
\end{eqnarray}
where the integration in eq. (\ref{RR}) is over all rotations $R$ and
$U_R$ is the  unitary transformation that realises rotation $R$.
Bob's task is thus to distinguish between the two density matrices
$\rho_\pm$. Since these density matrices are non orthogonal he only
has a finite chance of success. We will show below that this is a
general feature and that it is
impossible to make perfect rotationally invariant
 quantum gloves out of the relative
position of two particles. We expect that by increasing the size of
the Hilbert space, ie. by having the gloves have large angular
momentum, it is possible to make them better and better.

\subsection{Quantum gloves and decoherence free subspaces}

It is interesting to note that 
the quantum gloves constructed in the examples described above are
very closely related to the decoherence free subspaces
considered in \cite{ZM,LCW,BRS} and recently realized experimentally in
\cite{BEGKCW}. Indeed in these works the aim was to construct
orthogonal states
or subspaces that are invariant under rotation.
The main difference is that for these applications it is indifferent 
whether the subspaces are realized using spin
degrees of freedom or using relative position of particles. Thus for
instance the states realized in \cite{BEGKCW} are states of the
polarization (ie. angular momentum) 
of 4 photons and therefore are good decoherence free
spaces, but cannot serve as quantum gloves.

\section{Chirality operator}\label{VI}

In what preceeded we focused on the properties of the quantum states
$|G_\pm\rangle$ sent by Alice and supposed that Bob always measured
the same operator $\Chi$. This approach is the one which would be
adopted by the external observer if he has the same chirality as
Bob. In this section we consider the opposite situation where the
external observer has the same chirality as Alice. In this case Alice
always prepares the same state $|G_+\rangle$. But Bob will measure
either $\Chi$ or $P \Chi P$ according to his chirality. 

We study here 
the properties of the
chirality operator $\Chi$. This will provide us with a very general approach
to the problem of quantum gloves and will allow us to classify many
possible realizations of quantum gloves.
We will suppose that the quantum gloves are perfect, ie. that Bob can
perfectly distinguish whether or not Alice has the same chirality as
him. We will also suppose rotational invariance in the sense that we
require that a quantum glove $|G_+\rangle$ and the rotated glove
$R|G_+\rangle$ have exactly the same properties.

With these conditions the chirality operator must obey several conditions.
First of all the quantum gloves $|G_+\rangle$ and $|G_-\rangle$ must
be eigenstates of $\Chi$ with different eigenvalues:
\begin{eqnarray}
\Chi |G^+\rangle &=& \gamma^+ |G^+\rangle\nonumber\\
\Chi |G^-\rangle &=& \gamma^- |G^-\rangle\quad , \quad
\gamma^+ \neq \gamma^-
\label{C1}
\end{eqnarray}
with
\begin{eqnarray}
&\langle G^-|G^+\rangle =0\ ,&\label{C2A}\\
&P|G^+\rangle = |G^-\rangle \quad , \quad P|G^-\rangle = |G^+\rangle \
  .&\label{C2}
\end{eqnarray}
Properties (\ref{C1}) (or equivalently (\ref{C2A})) are necessary in
order to have perfect quantum gloves.

For simplicity we will suppose that the eigenvalues $\gamma^+$ and
$\gamma^-$ 
are opposite: $\gamma^- = - \gamma^+$ (although this is not essential
for the next part of the argument based on rotational invariance).
Then Eqs. (\ref{C1}) and (\ref{C2}) imply that
\begin{equation}
P \Chi P = - \Chi
\label{C3}
\end{equation}

Note that $\Chi$ may have a zero eigenvalue. The corresponding
eigenspace cannot be used to encode chirality. In what follows we
restrict our attention to the subspace on which $\Chi$ is non zero.

We now consider rotation invariance in the sense that we require that
if  $|G_+\rangle$ is a quantum glove then the rotated glove
$R|G_+\rangle$ have exactly the same properties. In particular this
implies that if $|G^+\rangle$ is an eigenstate of $\Chi$, 
then $R|G^+\rangle$ is also an eigenstate of $\Chi$ with the
same eigenvalue:
$$
\Chi R |G^+\rangle = \gamma^+ R |G^+\rangle\ .
$$
The same holds for $R|G_-\rangle$. 
These properties imply that the chirality operator is invariant under
rotation:
\begin{equation}
R^\dagger \Chi R = \Chi\quad \forall R \in SU_2 \ .
\label{C4} \end{equation}

Let us now consider the Hilbert space of the quantum gloves. For
definiteness we shall suppose that it is realized by some spin
degrees of freedom and the relative position of several particles. Let
us denote by $\vec L$ the total angular momentum operator acting on
this Hilbert space. Then
eq. (\ref{C4}) implies that $\Chi$ commutes with the generators of the
rotation group $\vec L$.
In particular $\Chi$ commutes with
the total angular momentum operator
$L^2 = L_x^2 + L_y^2 + L_z^2$. 

The addition properties of angular momentum imply that the Hilbert space 
decomposes into a direct sum of
spaces $H_L$ with total angular momentum $L$. We thus obtain that
$\Chi$ 
is block
diagonal in this representation. From now on we focus on a specific
subspace $H_L$ of total angular momentum $L$. 

The space of total angular
momentum $L$ can further be
decomposed into the direct sum of a number of irreducible
representations of $SU_2$. Some of these irreducible representations
have positive parity, denote them $H_{L+}$, whereas others have
negative parity, denote them $H_{L_-}$.
An arbitrary quantum glove of total angular momentum $L$ 
can thus be written as
$
|G^+_L\rangle =\alpha |\psi_{L+}\rangle + \beta |\psi_{L-}\rangle
$
where $|\psi_{L+}\rangle\in H_{L+}$ has total angular momentum L and
positive parity and $|\psi_{L-}\rangle\in H_{L-}$ has total angular
momentum L and 
negative parity.
Then the other glove has the form
$
|G^-_L\rangle =
P|G_+\rangle = \alpha |\psi_{L+}\rangle - \beta |\psi_{L-}\rangle$.
Orthogonality of the right and left gloves then implies that $\alpha
=\beta=1/\sqrt{2}$:
\begin{equation}
|G_\pm\rangle = {1\over \sqrt{2}} ( |\psi_{L+}\rangle \pm |\psi_{L-}\rangle)
\ .\label{C10}
\end{equation}

This provides a systematic way of constructing all possible quantum
gloves in a given Hilbert space. One simply decomposes the total
Hilbert space into a direct sum of spaces of different total angular
momentum and different parity. The quantum gloves are then arbitrary
states of the form eq. (\ref{C10}). The states constructed in the
previous sections  are particular examples of such quantum gloves
which use irreducible representations with small values of $L$. The
approach based on the chirality operator shows how to generalize this
to other values of $L$.

This also shows why one cannot construct rotationally invariant perfect quantum gloves using
the relative position of two particles. In this case to each value of
$L$ corresponds a single irreducible representation of $SU_2$ and one
cannot construct states of the form (\ref{C10}).

\section{conclusion}\label{VII}

We have shown that it is possible to construct
states, which we call ``quantum gloves'', that can be used to encode
the chirality of a reference frame. This is an apparently simple
question. But because the quantity one wants to encode is so simple
--it is only a dichotomic variable-- one can focus on the crucial 
role of the physical
properties of the system used to encode the information. We have seen
that whereas spin degrees of freedom cannot make quantum gloves,
relative positions of particles, or combinations of relative position
and spin, can make good quantum gloves. Furthermore one can make
tradeoffs between resources used: number of qubits transmitted versus
number of particles sent, versus volume occupied in space, etc...

In conclusion, quantum information can be thought of independently 
of any implementation, similarly to classical information. This 
rather trivial remark implies that quantum information can only 
achieve tasks which are expressed in pure information theoretical 
terms, like cloning and factoring, but can't perform physical 
tasks like aligning reference frames or defining temperature. Thus for
instance  
quantum teleportation is an information concept and 
does not permit the teleportation of a physical object, including 
its mass and chirality. This underlines that {\it information is 
physical}, but {\it physics is more than just information}.

{\bf Acknowledgements:} The authors acknowledge financial support from
the European Union
through project RESQ IST-2001-37559 and by the swiss NCCR Quantum 
Photonics. S.M. acknowleges support by the
Action de Recherche Concert{\'e}e de la Communaut\'e Fran{\c{c}}aise de
Belgique and by the IUAP program of the Belgian Federal Governement
under grant V-18. S.M. would like to thank Michel Tytgat for helpful
discussions.

{\bf APPENDIX: CHIRALITY AND PARTICLE PHYSICS}

In the introduction we noted that ``if our world is 
invariant under {\it left} $\leftrightarrow$ {\it right}, then 
mere information is unable to distinguish between {\it left} and 
{\it right}''. But of course particle physics has tought us that our
world is not invariant under  {\it left} $\leftrightarrow$ {\it
  right}. Indeed the Hamiltonian of elementary particle physics,
describing the behavior of kaons, etc... is not invariant under {\it
  left} $\leftrightarrow$ {\it right}.
Thus one can prepare a quantum state of elementary particles
$|\psi_0\rangle$ which is invariant under parity, $P|\psi_0\rangle =
|\psi_0\rangle$, let it evolve, and the final state $e^{-i H t}
|\psi_0\rangle$ is no longer invariant under parity.
Thus the universe is in fact endowed with an absolute chirality. In
this case Alice no longer needs to reveal to Bob some physical
information. She only needs to measure her chirality with respect to
the absolute chirality of the universe and tell the result to
Bob. This can be done using information only, ie. using only black and white
balls. See \cite{F} for a discussion.

In the present work we have 
supposed that Alice and Bob do not have access
to a parity violating Hamiltonian and are restricted to manipulating
some simple degrees of freedom such as spin, position,
etc... 
In this case the chirality of their reference frame must be
encoded in the quantum states they use. Thus the question we study here
is: what are the physical degrees of freedom that allow 
one to encode chirality, and what is the most economical way of doing
so, if one does not have access to a parity violating Hamiltonian.
It would
certainly be very 
interesting to revisit this problem in the light of the known
properties of the particle 
physics Hamiltonian. For instance some particles, such as pions, have
an intrinsic parity, and this could presumably be exploited when
constructing quantum gloves.

\end{document}